\documentclass[sigconf]{acmart}

\AtBeginDocument{%
  }

\copyrightyear{2024}
\acmYear{2024}
\setcopyright{acmlicensed}
\acmConference[x 'XX]{x}{July 03--05,
  2024}{Glasgow, UK}
\acmDOI{XXXXXXX.XXXXXXX}

\usepackage{microtype}
\usepackage{dsfont}
\usepackage[para]{footmisc}
\usepackage[show]{chato-notes}
\usepackage{enumitem}
\usepackage{hyperref}
\usepackage{booktabs}  
\usepackage{graphicx}  
\usepackage{makecell}  
\usepackage{array}
\usepackage[normalem]{ulem}
\usepackage{multirow}
\usepackage{balance}  
\newcommand{\pageenlarge}[1]{\enlargethispage{#1\baselineskip}}

\newcommand{\zm}[1]{\textcolor{black}{#1}}

\newcommand{\qq}[1]{\textcolor{black}{#1}}
\newcommand{\zym}[1]{\textcolor{black}{#1}}

\definecolor{darkblue}{rgb}{0.0, 0.0, 0.55}
\begin{document}
\title{RecRankerEval: A Flexible and Extensible Framework for Top-k LLM-based Recommendation}

\author{Zeyuan Meng}
\affiliation{%
  \institution{University of Glasgow}
  \city{Glasgow}
  \country{UK}}
  \email{z.meng.2@research.gla.ac.uk}

\author{Zixuan Yi}
\affiliation{%
  \institution{University of Glasgow}
  \city{Glasgow}
  \country{UK}}
  \email{z.yi.1@research.gla.ac.uk}

\author{Iadh Ounis}
\affiliation{%
  \institution{University of Glasgow}
  \city{Glasgow}
  \country{UK}}
  \email{iadh.ounis@glasgow.gla.ac.uk}

\renewcommand{\shortauthors}{Meng et al.}

\begin{abstract}
A recent Large language model (LLM)-based recommendation model, called RecRanker, has demonstrated a superior performance in the top-K recommendation task compared to other models. 
In particular, RecRanker samples users via clustering, generates an initial ranking list using an initial recommendation model, and fine-tunes an LLM through hybrid instruction tuning (listwise, pairwise, pointwise) to infer user preferences. 
However, the contribution of each core component—user sampling, initial ranking list generation, prompt construction, and hybrid instruction tuning strategy—remains underexplored. 
In this work, we inspect the reproducibility of RecRanker, and study the impact and role of its various components. 
We begin by reproducing the RecRanker pipeline through the implementation of all its key components. 
Our reproduction shows that the pairwise and listwise methods achieve a performance comparable to that reported in the original paper. 
For the pointwise method, while we are also able to reproduce the original paper’s results, further analysis shows that the performance is abnormally high due to data leakage from the inclusion of ground-truth information in the prompts. 
To enable a fair and comprehensive evaluation of LLM-based top-$k$ recommendations, we propose RecRankerEval—an extensible framework that covers five key dimensions: user sampling strategy, initial recommendation model, LLM backbone, dataset selection, and instruction tuning method.
Using the RecRankerEval framework, we show that the original results of RecRanker can be reproduced on the ML-100K and ML-1M datasets, as well as the additional Amazon-Music dataset, but not on BookCrossing due to the lack of timestamp information in the original RecRanker paper. 
Furthermore, we demonstrate that RecRanker's performance can be improved by employing alternative user sampling methods, stronger initial recommenders, and more capable LLMs. 
Our RecRankerEval framework offers a flexible platform for deploying and evaluating various configurations of RecRanker, paving the way for a further in-depth analysis of LLM-based recommenders.

\looseness -1
\pageenlarge{2}
\end{abstract}



\maketitle

\section{Introduction}\label{s1}
\pageenlarge{2}
Large Language Models (LLMs)~\cite{shi2024large,zhang2024lorec,sun2024large} have recently gained significant attention in the field of recommender systems (RS) for their strong reasoning capabilities and \qq{their} ability to model user intent through natural language.
Existing LLM-based recommendation models~\cite{li2024calrec,lin2024data} typically formulate recommendations as a prompt-based task, allowing LLMs to infer and reason about user preferences, thereby improving \qq{the} recommendation performance.
To better align LLMs with \qq{the} recommendation tasks, these models are often fine-tuned on domain-specific datasets through instruction tuning—a technique that uses task-specific prompts to guide the model during training~\cite{li2023exploring,wu2024survey,yi2023contrastive}.
A prominent example is RecRanker~\cite{luo2023recranker}, one of the first LLM-based recommendation models to apply instruction tuning for top-$k$ recommendation.
Specifically, RecRanker first applies clustering-based user sampling to select representative users for prompt construction. 
It then employs \zym{an} initial recommendation model
to generate \zym{an} initial ranking 
\zym{list,} enriching the 
\zym{prompts with relevant user–item contexts.}
Finally, it fine-tunes the LLM using a hybrid of three instruction tuning strategies—pointwise, pairwise, and listwise—to improve its ability to generate accurate top-$k$ recommendation lists.
Given its early influence, novel design, and reported effectiveness, RecRanker serves as a strong foundation for LLM-based recommenders.
However, reproducing RecRanker’s results remains challenging, as the released repository lacks essential data processing scripts and training code, preventing a full replication of its experimental pipeline.
Furthermore, the impact of RecRanker’s core components—user sampling,  initial ranking list generation, prompt construction, and hybrid instruction tuning strategy—on \qq{the} recommendation performance has not been thoroughly investigated \qq{in previous studies}.
To address these gaps, we begin by reproducing the RecRanker model to assess whether its reported findings hold under reimplementation.
Our reproduction reveals both consistent results and unexpected anomalies, motivating a deeper investigation into RecRanker's design.
To facilitate this, we \qq{introduce} RecRankerEval, a flexible evaluation framework that supports systematic analysis across five key dimensions of \qq{LLM-based} top-$k$ \qq{recommendations}: user sampling strategy, initial recommendation model, datasets selection, LLM backbone, and instruction tuning method.
\looseness -1

In this work, our primary contributions are as follows:
(1) We reimplement all major components of the RecRanker 
\zym{pipeline},
including user sampling, initial ranking list generation, prompt construction, and hybrid instruction tuning strategy.
Our reproduction confirms the 
\zym{effectiveness}
of the listwise and pairwise methods as reported in the original paper.
However, although the pointwise results are reproducible, further analysis reveals that the \qq{obtained} high performance results are due to data leakage\zym{;}
(2) We propose RecRankerEval, an extensible framework 
\zym{that builds upon RecRanker by enabling flexible substitutions and extensions \qq{of} each component.}
\zym{This framework allows us to explore improved configurations and to gain deeper insights into which design choices contribute most to recommendation performance;}
(3) 
\zym{We \qq{analyse} the generalisability of RecRankerEval by extending \qq{the} evaluation to a new domain, 
\zym{namely}
the Amazon-Music dataset from the Amazon review corpus.}
\looseness -1

The \qq{remaining sections of the paper are organised} as follows:
Section~\ref{s2} 
\zym{presents}
related work on LLM-based recommender systems, with a particular focus on the modelling choices 
\zym{used}
in RecRanker.
\zym{Section~\ref{s3} describes the RecRanker model in detail, laying the foundation for our reproduction and component-wise analysis.}
Section~\ref{s4} introduces our proposed framework, RecRankerEval, 
\zym{an extensible evaluation framework that not only reproduces RecRanker’s pipeline, but also supports systematic comparison across alternative configurations of its five core components.}
Section~\ref{s5} presents our reproduced results of RecRanker, 
\zym{identifies the potential data leakage issue, and systematically evaluates the impact of different components using RecRankerEval across multiple datasets.}
Finally, \qq{Section~\ref{s6} provides} 
\zym{the conclusions of our work.}

\section{Related Work}\label{s2}
\pageenlarge{3}
Recently, Large Language Models (LLMs) have been increasingly explored for their potential in recommender systems, due to their strong contextual reasoning and generalisation capabilities~\cite{lin2024data,li2024calrec,yi2025enhancing,yi2025multi,yi2023large}.
Two main paradigms have emerged for applying LLMs to recommendation tasks\zym{--}non-tuning and tuning\zym{--}
which differ in how they adapt pretrained models to downstream \qq{tasks} use~\cite{wu2024survey,zhao2024recommender,yi2022multi,yi2024directional,yi2023graph,meng2024knowledge,yi2024unified,meng2024knowledge,yi2025multi}.
The non-tuning paradigm leverages the zero- and few-shot reasoning capabilities of LLMs by 
\zym{constructing}
prompts that encode user–item interactions.
For instance, Zhang et al.~\cite{zhang2021language} used a pretrained model for conversational recommendation without any task-specific training.
\zym{However, due to the limited input token length of LLMs, zero-shot prompts can only include a small number of items, which may negatively impact \qq{the} recommendation performance~\cite{geng2022recommendation,shin2021one4all}.}
To address this limitation, 
\zym{later LLM-based recommendation models proposed 
\zym{identifying}
the most relevant textual information of items and users \zym{as} input to \qq{the} LLMs, thereby improving the recommendation performance~\cite{cui2022m6}.}
\zym{For example, NIR~\cite{wang2023zero} first identifies candidate items that are most relevant to the user based on interaction history or item features, effectively reducing the input space, and then applies a three-stage prompting strategy for sequential recommendation.}
Few-shot methods further improve performance by including demonstration examples in the prompts~\cite{wang2024empowering,zhang2024label}.
Dai et al.~\cite{dai2023uncovering} found that using \zym{demonstration} templates based on pointwise, pairwise, and listwise ranking strategies achieved better results than zero-shot prompts.
\zym{Despite these advances, few-shot methods struggle to align LLMs with the specific objectives of recommendation tasks, due to their lack of task-specific supervision~\cite{petrov2023generative,wang2023zero}.}
To bridge this gap, tuning-based methods fine-tune LLMs using natural language instructions, allowing models to learn domain-specific patterns~\cite{bao2023tallrec,ji2024genrec}.
InstructRec~\cite{zhang2023recommendation} \zym{was} the first \qq{model} to adopt this approach, designing instruction formats for tasks such as sequential recommendation and product search.
However, existing tuning methods often rely on randomly sampled interactions and a single tuning strategy, limiting their ability to fully exploit LLMs’ potential\zym{~\cite{geng2022recommendation,wei2024llmrec}}.
\zym{Unlike prior tuning-based methods,}
RecRanker~\cite{luo2023recranker} 
\zym{addresses these limitations and}
is the first LLM-based framework explicitly designed for the top-$k$ recommendation task.
It addresses 
\zym{the limitations of random user sampling and single tuning strategy}
by using user sampling to construct three instruction tuning prompts—listwise, pairwise, and pointwise—and fine-tuning Llama2~\cite{touvron2023llama} accordingly, which enhances alignment 
\zym{between the LLM and} the top-$k$ recommendation task, \zym{thereby improving} performance.\looseness-1

\zym{Despite its promising design, RecRanker still \qq{presents} critical reproducibility challenges.}
\zym{The original implementation lacks essential data processing and training scripts, and its instruction tuning relies on full fine-tuning with 16 NVIDIA A800 80GB GPUs—posing a significant barrier for resource-constrained environments. 
Moreover, the original paper provides limited evaluation of its \zym{four} core components, 
leaving open questions about their individual impact on performance.}
\zym{To address these limitations, we reimplement RecRanker and introduce RecRankerEval—a flexible evaluation framework that supports controlled reproduction and targeted analysis.}
\zym{Rather than treating RecRanker as a monolithic design, RecRankerEval enables systematic inspection of its architecture, offering insights into design effectiveness and opportunities for further optimisation.}

\section{RecRanker Model}\label{s3}
\pageenlarge{3}
To support our reproduction study and examine the effectiveness of LLM-based top-$k$ recommendation, we first focus on RecRanker, the first LLM-based model explicitly designed for this task.
RecRanker has gained increasing attention for its novel integration of LLMs with structured user sampling and prompt design, yet its implementation details and component-level contributions remain underexplored.
RecRanker includes four core components—user sampling, 
initial ranking list generation, prompt construction, and hybrid instruction tuning strategy.
\zym{These components work together to reformulate the top-$k$ recommendation task into a natural language format, making it suitable for \qq{processing by} large language models.}
In the following subsections \zym{(Sections~\ref{s3.1}–~\ref{s3.4})}, we describe each component \qq{of RecRanker} as the foundation for our reproduction and extension in RecRankerEval.
\looseness -1

\subsection{User Sampling}\label{s3.1}
User sampling in RecRanker aims to select \qq{a} representative and diverse \zym{set of} users for constructing high-quality prompts to fine-tune the LLM.
\zym{To achieve this,}
RecRanker employs three strategies for user sampling:
\noindent \textbf{(1) Importance-aware sampling}: 
\zym{RecRanker} prioritises users with more interactions by assigning each user $u$ a sampling probability proportional to their interaction-based importance score. 
Specifically, the probability of selecting user $u$ is given by:
$p_{u,\text{importance}}=\frac{\ln(q_u)}{\sum_{v\in\mathcal{U}}\ln(q_v)}$,
where $q_u$ denotes the number of interactions for user $u$, 
\zym{$\mathcal{U}$ is the full user set, and $\ln$ represents the natural logarithm.}
The resulting sampled set is denoted as $\mathcal{U}_1\sim p_{u,\text{importance}}$;
\noindent \textbf{(2) Clustering-based sampling}:
\zym{RecRanker applies K-means~\cite{hartigan1979algorithm} to cluster users based on their interaction embeddings, then samples users proportionally from each cluster to ensure diversity across intent groups.}
Each user $u$ is assigned to a cluster $k_u \in \{1, \ldots, K\}$,
and the sampling probability is set proportional to the cluster size, 
i.e., $p_{u,\text{clustering}} \propto \left| \{v \in \mathcal{U} : k_v = k_u \} \right|$.
The sampled user set is denoted as $\mathcal{U}_2\sim p_{u,\text{clustering}}$;
\noindent \textbf{(3) Penalty for repeated selection}:
reduces repeated sampling of the same users.
Given the merged multiset $\mathcal{U}_3 = \mathcal{U}_1 \cup \mathcal{U}_2
$,
this strategy introduces a penalty weight
$\psi_u = C^{M_3(u)}$,
where $C \in (0, 1)$ is a constant and $M_3(u)$ is the number of times user $u$ appears in $\mathcal{U}_3$. 
The probability of selecting user $u$ is denoted as
$\mathcal{U}_3\sim p_{u,\text{penalty}}$.  
\zym{Together, these \zym{three} sampling methods ensure that the selected users provide diverse and informative training signals \zym{to the LLM}, forming a solid foundation for subsequent prompt construction and instruction tuning.}
\looseness -1

\subsection{\zym{Initial Ranking List Generation}}\label{s3.2}
The second component of RecRanker is the initial ranking list generation, which differs between the training and inference phases.
During training, RecRanker forms a training ranking list for each user $u$ by combining interacted items ($I_u^+$) and negatively sampled items ($I_u^-$), drawn from the set of unobserved items $\mathcal{I} \setminus I_u$:
$R_u^{\text{train}} = I_u^+ \cup I_u^-$.
Here, $R_u^{\text{train}}$ typically includes 10 items, with negative samples incorporated based on the assumption that unobserved items are more likely to be preferred than those explicitly disliked. 
\zym{This strategy enables the model to distinguish between relevant and \qq{non-relevant} items during training.}
During inference, RecRanker uses an initial recommendation model—such as MF~\cite{koren2009matrix} or LightGCN~\cite{he2020lightgcn}—to assign scores to all unseen items using a prediction function $f_{\text{init}}(u, i)$. The top-$k$ items are selected to construct the inference ranking list:
$R_u^{\text{infer}} = \text{TopK}_{i \in \mathcal{I} \setminus I_u} f_{\text{init}}(u, i)$.
In both phases, RecRanker 
\zym{typically}
selects a ranking list of 10 items as input to the subsequent prompt construction component.

\subsection{Prompt Construction}\label{s3.3}
The third component of RecRanker is prompt construction, 
\zym{which generates natural language inputs to guide the LLM during instruction tuning and inference.}
RecRanker 
\zym{devises}
three ranking prompt types—listwise, pointwise, and pairwise—aimed at translating the recommendation task into a format suitable for language model processing.
As shown in Table~\ref{tab1}, each prompt consists of four parts: 
\noindent \textbf{(1) Task description}: specifies that the goal is to perform a top-$k$ recommendation \qq{task};
\noindent \textbf{(2) User interaction history}: includes both liked and disliked items. During training, interactions are selected through user sampling; during inference, they are randomly sampled from each user’s 
\zym{historical interactions.}
\noindent \textbf{(3) Ranking context}: 
all three prompt types use the top-$k$ results from the initial recommendation model, but format these items differently using method-specific natural language templates
\zym{(see Table~\ref{tab1} for examples).}
\noindent \textbf{(4) \zym{Initial ranking hints}}: 
\zym{each prompt incorporates information from the initial ranking list generated by the recommendation model—such as predicted item ranks or scores—to guide the LLM’s ranking decisions.}
\begin{table}
\centering
\scriptsize  
\setlength{\tabcolsep}{3pt}
\caption{Illustrative examples of instructions for three ranking methods.} 
\begin{tabular}{|l|p{7.2cm}|}
\hline
\textbf{Type} & \textbf{Instructions}\\ \hline
\textbf{Listwise} & You are a movie recommender system. Your task is to rank a given list of candidate movies based on user preferences and return the top five recommendations. 
\newline User's Liked movies: \underline{\textless  liked historical interactions\textgreater}. 
\newline User's Disliked movies: \underline{\textless disliked historical interactions\textgreater}.
\newline Question: How would the user rank the candidate item list: \underline{\textless ranking item list\textgreater}?
\newline Hint: Another recommender model suggests \underline{\textless ranking list\textgreater}.
\\ \hline
\textbf{Pointwise} & You are a movie recommender system. Your task is to predict the relevance score to a target movie based on the user's historical movie ratings. 
\newline The score should be between 1 and 5. 
\newline User's Liked movies: \underline{\textless liked historical interactions\textgreater}. 
\newline User's Disliked movies: \underline{\textless disliked historical interactions\textgreater}.
\newline Question: Based on the user's historical ratings, predict the relevance score of the target <item> with the user.
\newline Hint: Another recommender model suggests the answer is \underline{\textless score\textgreater}.
 \\ \hline
\textbf{Pairwise} & You are a movie recommender system. Based on a user's likes and dislikes, determine if they would prefer one movie over another. Respond only with "Yes." or "No.". 
\newline User's Liked movies: \underline{\textless  liked historical interactions\textgreater}. 
\newline User's Disliked movies: \underline{\textless disliked historical interactions\textgreater}.
\newline Question: Would the user prefer <movie1> over <movie2>?
\newline Hint: Another recommender model suggests the answer is \underline{\textless movie1\textgreater}.
 \\ \hline
\end{tabular}
\label{tab1}
\vspace{-6mm}
\end{table}

\subsection{\zym{Hybrid Instruction Tuning Strategy}}\label{s3.4}
\pageenlarge{3}
The final component of RecRanker is \qq{the} hybrid instruction tuning strategy.
RecRanker performs full fine-tuning by feeding prompts into Llama2, aligning the LLM with the top-$k$ recommendation task. 
\zym{Specifically, RecRanker fine-tunes the pretrained LLM by minimising the cross-entropy loss~\cite{zhang2018generalized}.}
This tuning process enhances the LLM’s ability to capture user intent and preference patterns more effectively.
\zym{After tuning, the LLM uses the three prompt types—pointwise, pairwise, and listwise—for inference, generating individual recommendation outputs for \qq{the} target users.}
\zym{RecRanker then applies a hybrid ranking strategy that integrates the outputs from the pointwise, pairwise, and listwise prompts to produce the final top-$k$ recommendation list.}
In the pointwise setting, each item’s utility score is computed as: $U_{pointwise}= P + U_{retrieval}$, where $U_{retrieval}=-m\cdot C_1$, with $P$ being the LLM-predicted relevance score, $m$ the item’s rank in the initial recommender output, and $C_1$ \qq{is} a predefined constant.
\zm{In the pairwise setting, each item receives a constant utility score: $U_{pairwise} = C_2$, where $C_2$ is a constant.}
\zm{In the listwise setting, the utility score depends on the item’s position in the LLM-generated ranking: $U_{listwise} = -m' \cdot C_3$, where $m'$ is the rank assigned by the LLM and $C_3$ is a constant.}
\zm{RecRanker computes the final hybrid score for each item by linearly combining the above scores:}
$U_{hybrid} = \alpha_1 U_{pointwise} + \alpha_2 U_{pairwise} + \alpha_3 U_{listwise}$,
\zm{where $\alpha_1 + \alpha_2 + \alpha_3 = 1$.}

\zm{RecRanker is the first model to integrate LLMs into top-$k$ recommendation, but \qq{as previously mentioned} reproducing its results presents three challenges:}\looseness -1

\noindent $\bullet$ 
The RecRanker codebase lacks both processed datasets and the necessary data processing scripts.
\zm{It also omits critical scripts for generating intermediate files—such as \qq{the} user embedding 
\zym{files}
and item prediction score 
\zym{files}—required for prompt construction.}
\zm{This gap makes it difficult to \qq{reproduce} RecRanker’s third component;}\looseness -1

\noindent $\bullet$ 
For the fourth component, RecRanker adopts full fine-tuning on Llama2 using 16 NVIDIA A800 80GB GPUs. 
This setup requires substantial computational resources, making the original 
\zym{RecRanker}
training stage impractical for resource-constrained environments; \looseness -1

\noindent $\bullet$ 
\zm{The original RecRanker paper lacks sufficient ablation study or component-level analysis of its four key modules--user sampling, initial ranking list generation, prompt construction, and hybrid instruction tuning strategy--leaving their individual contributions to \qq{the} overall performance unclear.}
To address these challenges, we reimplement RecRanker within our RecRankerEval framework and verify its reported performance.
Beyond reproduction, RecRankerEval enables a deeper investigation into the model’s design decisions by supporting controlled comparisons, alternative configurations, and targeted analysis of potential issues-such as component sensitivity and data leakage\zym{-as discussed in Sections~\ref{s5.4} to~\ref{s5.7}.}
\begin{figure*}[t]
\centering
\includegraphics[width=0.8\linewidth]{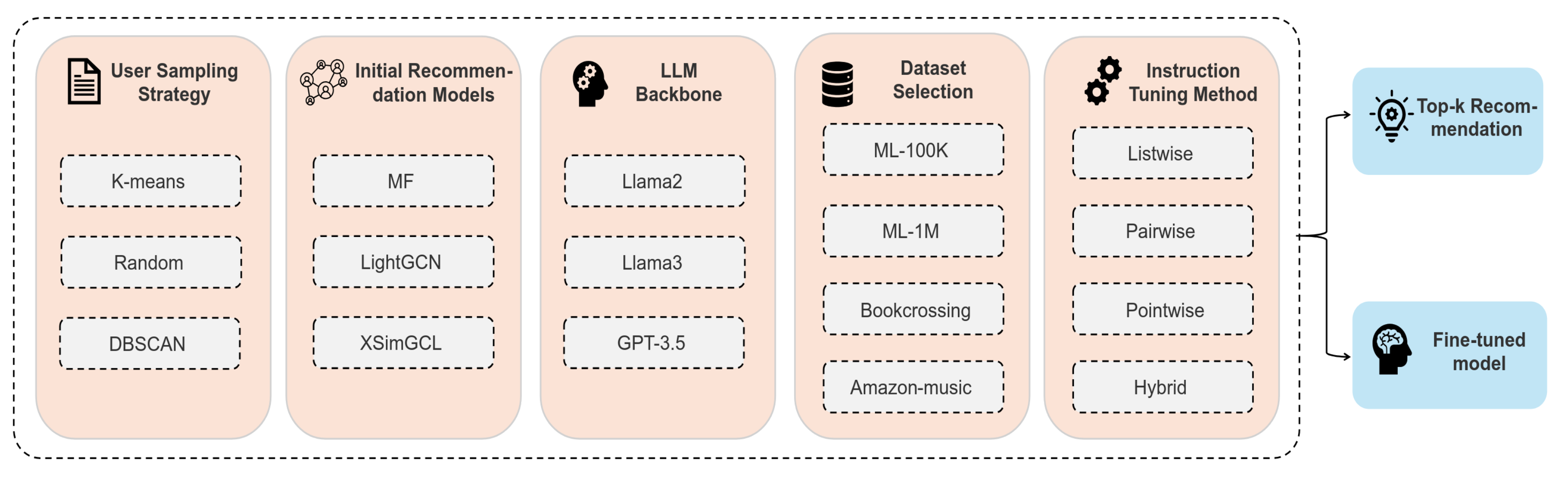}
\vspace{-6mm}
\caption{\qq{Overview of the} RecRankerEval Framework.} 

\label{fig1}\vspace{-2mm}
\end{figure*}
\section{RecRankerEval}\label{s4}
\pageenlarge{3}
\zym{In this section, we start by providing an overview of the RecRankerEval framework (Section~\ref{s4.1}). \qq{Next,} 
Sections~\ref{s4.2} to~\ref{s4.6} introduce the five core components of the framework: user sampling strategy, initial recommendation model, LLM backbone, dataset selection, and instruction tuning method.}
\subsection{\zym{Framework Overview}}\label{s4.1}
\qq{RecRankerEval is} an extensible framework built upon RecRanker to enable \qq{the} reproducible and systematic evaluation of LLM-based top-$k$ recommendation models.
\qq{One key objective of} RecRankerEval \qq{is} to support modular experimentation and \qq{a} comparative analysis \qq{of different alternative choices when deploying an LLM-based recommender such as RecRanker. Such choices pertain to five key components: user sampling strategy, initial recommendation model, LLM backbone, dataset selection, and instruction tuning method}. 
\zym{As illustrated in Figure~\ref{fig1}, RecRankerEval offers standardised interfaces, evaluation protocols, and support for diverse configuration options.}
\zym{The current \qq{framework} implementation includes three user sampling strategies (random, K-means, DBSCAN), three initial recommendation models (MF, LightGCN, XSimGCL), three LLM backbones (Llama2, Llama3, GPT-3.5), four datasets (ML-100K, ML-1M, BookCrossing, Amazon-Music), and four instruction tuning methods (pointwise, pairwise, listwise, hybrid).}
Beyond these built-in modules, the framework is designed for extensibility—allowing researchers to integrate new models, datasets, and LLMs with minimal effort. 
This makes RecRankerEval a generalisable platform for controlled experimentation and in-depth analysis of LLM-based top‑$k$ systems 
\zym{that follow or extend the RecRanker paradigm.}
\looseness -1
\subsection{User Sampling Strategy}\label{s4.2}
\pageenlarge{4}
\zym{To reproduce the user sampling component described in Section~\ref{s3.1}, RecRankerEval implements all three strategies used in the original RecRanker: importance-aware sampling, clustering-based sampling, and repetition penalty.}
\zym{While retaining the importance-aware and repetition penalty mechanisms, we explore alternatives specifically for the clustering-based method, which originally used K-means~\cite{luo2023recranker}.}
\qq{Indeed, in addition} \zym{to K-means}, our framework \zym{also} introduces two alternative sampling strategies—random sampling and DBSCAN (Density-Based Spatial Clustering of Applications with Noise)~\cite{wang2015adaptive} sampling—to enable \qq{for a} broader investigation into how different user selection mechanisms impact final recommendation performance.
For random sampling, we randomly select a subset of users for training:
$U_{\text{random}} \sim \text{Uniform}(\mathcal{U})$,
where $\mathcal{U}$ is the full user set.
For DBSCAN-based sampling, we cluster user embeddings and perform denoised sampling by excluding noise points and selecting users proportionally from valid clusters.
The probability of selecting a user $u$ is proportional to the size of their cluster:
$p_{u,\text{DBSCAN}} \propto |\{v \in \mathcal{U} : k_v = k_u,\, k_u \ne -1\}|$,
where $k_u$ denotes the cluster label assigned to user $u$ by DBSCAN, and $k_u = -1$ indicates a noise point. 
\zym{This clustering-based strategy helps reduce the influence of outliers by discarding users labelled as noise ($k_u = -1$), which are typically far from any dense user group in the embedding space.}
\zym{
When replacing K-means with \qq{the} DBSCAN-based sampling, we keep the remaining components of RecRanker's original three-part sampling strategy unchanged. 
\zym{In contrast, the random sampling strategy is used as a standalone alternative, where training users are selected uniformly at random without incorporating importance or repetition considerations.}
This setup enables a systematic investigation into how alternative clustering methods affect the diversity and representativeness of \qq{the} sampled users, and their downstream impact on LLM-based top-$k$ recommendation performance.}
\looseness -1
\subsection{Initial Recommendation Model}\label{s4.3}
\pageenlarge{4}
\zym{RecRankerEval reproduces the second component of the original RecRanker pipeline by \qq{providing} initial recommendation models that generate an initial ranking list for each user. 
This ranking list serves as the basis for constructing prompts, enabling the LLM to operate within a user-specific recommendation context during instruction tuning.}
\zm{As shown in Figure~\ref{fig1},}
\zym{our \qq{current} RecRankerEval implementation includes}
three initial recommendation models: Matrix Factorisation (MF)\zym{~\cite{koren2009matrix}} and LightGCN\zym{~\cite{he2020lightgcn}}, 
\zym{both} used in the original RecRanker paper, and XSimGCL\zym{~\cite{yu2023xsimgcl}}, 
a recent self-supervised graph-based model shown to improve recommendation accuracy under sparse interactions. 
We include XSimGCL to expand the evaluation scope and test RecRankerEval under more advanced 
\zym{initial recommendation model settings.}
\qq{This allows to examine} how different initial recommendation models 
\zym{affect the initial ranking lists used in \qq{the} prompt construction, \qq{thus} indirectly influencing the effectiveness of subsequent instruction tuning and \qq{the} final recommendation performance.}
\zym{To ensure reproducibility, we implement the complete pipeline for each initial recommendation model, including the generation of intermediate outputs such as \qq{the} user embeddings and \qq{the} ranked item lists, which are essential for prompt construction but were not released in the original RecRanker implementation.}
\qq{It is worth noting that additional recommendation models can be seamlessly integrated into RecRankerEval with minimal effort.}
\looseness -1
\subsection{Large Language Model (LLM) Backbone}\label{s4.4}
To systematically investigate the role of LLMs in top-$k$ recommendation, our RecRankerEval framework supports both instruction-tuned and zero-shot settings, covering a range of model types and usage modes.
Specifically, we categorise LLM integration into two paradigms: instruction-tuned LLMs, which are 
\zym{open}
models capable of being fine-tuned on prompt data to better align with the recommendation task; and zero-shot LLMs, which are 
\zym{closed}
models that cannot be fine-tuned and are instead used directly via prompting.
\zm{As shown in Figure~\ref{fig1},}
RecRankerEval supports Llama2 and Llama3 as 
\zym{open}
models, both of which are fully integrated into the \zym{RecRankerEval} training pipeline 
and used in the instruction-tuning setting. 
Our framework also incorporates GPT-3.5 as a 
\zym{closed}
model, 
\zym{and applies it in a zero-shot setting, since it is only accessible via API and does not support fine-tuning.}
\zm{By supporting these diverse configurations \qq{in} RecRankerEval, \qq{we can explore} how \qq{the top-$k$ recommendation} performance varies across \qq{different LLMs} (open vs. closed), adaptability \qq{settings} (tunable vs. non-tunable), and application \qq{modes} (instruction-tuned vs. zero-shot).} 
\looseness -1
\subsection{\zym{Dataset Selection}}\label{s4.5}
\pageenlarge{4}
As shown in Figure~\ref{fig1}, we evaluate RecRanker \qq{and its variants} on four widely used public datasets.
\zym{\qq{Results on the} ML-100K and ML-1M are directly reproduced from the original RecRanker's 
\zym{supplied resources}~\cite{luo2023recranker}; \qq{Results on the BookCrossing dataset are reproduced} based on the descriptions in \qq{the RecRanker's} original paper~\cite{ziegler2005improving}; \qq{while we also experiment with the additional} Amazon-Music \qq{dataset} to assess the generalisability of the 
\zym{RecRanker and its variants}
across \qq{different} domains.}
\qq{More specifically, while for} ML-100K and ML-1M, we reproduce the dataset configurations based on the original RecRanker paper~\cite{luo2023recranker}, 
since the original implementation does not provide processing scripts for BookCrossing, we \qq{reproduce} the full data preparation pipeline according to the descriptions in the paper.
In addition, we \qq{also similarly process the additional} 
\zym{Amazon-Music dataset}.
To ensure reproducibility, we provide complete data processing scripts for 
\zym{all datasets,}
allowing researchers to recreate the experimental datasets from raw sources. 
Detailed preprocessing \zym{descriptions} for all datasets are provided in Section~\ref{s5.1}.\looseness -1
\subsection{Instruction Tuning Method}\label{s4.6}
As shown in Figure~\ref{fig1}, our RecRankerEval framework implements \qq{the} four instruction tuning methods
\zym{introduced in \qq{the original} RecRanker \qq{paper namely the three}, pointwise, pairwise and listwise \qq{methods as well as} a hybrid strategy that combines all three.}
For each 
\zym{method, we fine-tune the} {open LLMs}—specifically Llama2 and Llama3—using prompts 
\zym{that correspond to the method’s ranking objective as described in Section~\ref{s3.3}.}
\zym{These prompts are constructed to match the ranking objective}
of each 
\zym{instruction tuning method}
and guide the LLM in learning to rank items accordingly.
Once fine-tuned, the 
\zym{LLMs}
generate top-$k$ recommendation lists based on the given prompts.
\zm{To achieve \qq{a} fair comparison across \qq{the various} tuning 
\zym{methods,}
RecRankerEval provides standardised output formats and evaluation templates. 
This addresses a key limitation of the original RecRanker implementation, which did not release standardised output processing scripts or formatting conventions.}
\zm{In addition, we provide scripts that automatically process the raw model outputs—such as item-score pairs or ranked lists—into final recommendation results suitable for evaluation. 
This automation streamlines the pipeline and ensures \qq{a} consistent evaluation across different instruction tuning setups.}
\looseness -1

\section{Experiments}\label{s5}
\zym{In Sections~\ref{s5.1} and ~\ref{s5.2},}
we describe the datasets and baseline models used throughout our experiments. 
Section~\ref{s5.3} outlines the evaluation metrics and provides implementation details \qq{about} 
\zym{our} RecRankerEval framework. 
Sections~\ref{s5.4} to~\ref{s5.7} 
\zym{present} and analyse \zym{the} experimental results, each addressing a specific research question.
We organise our analysis around 
\zym{four}
research questions:

\noindent \textbf{RQ1}: 
\zym{Can we reproduce the reported results of RecRanker \qq{both} on \qq{the originally} used and additional datasets using our RecRankerEval framework?}\looseness -1

\noindent \textbf{RQ2}: 
How do different initial recommendation models affect the performance of RecRankerEval in top-$k$ recommendation?

\noindent \textbf{RQ3}: 
\zym{How do different user sampling strategies affect the performance of RecRankerEval in \zym{the} top-$k$ recommendation \qq{task}?}

\noindent \textbf{RQ4}: 
How does the \zym{choice of} \qq{the} LLM \zym{backbone} influence the performance of RecRankerEval across \qq{the} zero-shot and instruction tuning paradigms?



\vspace{-2mm}
\subsection{\zym{Datasets}}\label{s5.1}
\pageenlarge{4}
\begin{table}
\caption{Statistics of the used datasets.}
\label{tab:1}
\centering
\resizebox{1.0\linewidth}{!}{ 
\begin{tabular}{c|llll|llll}
    \hline
    \multirow{2}{*}{Dataset} & \multicolumn{4}{c|}{Original Paper} & \multicolumn{4}{c}{Reproduced} \\
    \cline{2-9}
    & Users & Items & Interactions & Density & Users & Items & Interactions & Density \\
    \hline
    ML-100K & 943 & 1,682 & 100,000 & 0.063046 & 943 & 1,682 & 100,000 & 0.063046 \\
    ML-1M & 6,040 & 3,706 & 1,000,209 & 0.044683 & 6,040 & 3,952 & 1,000,209 & 0.041902 \\
    BookCrossing & 1,820 & 2,030 & 41,456 & 0.011220 & 1,777 & 22,288 & 90,818 & 0.002293 \\
    Amazon-Music & / & / & / & / & 1,162 & 3,935 & 30,642 & 0.006701 \\
    \hline
\end{tabular}
}
\label{tab2}
\end{table}
\zym{As introduced in Section~\ref{s4.5}, \qq{we conduct} experiments on four datasets: MovieLens-100K (ML-100K), MovieLens-1M (ML-1M), BookCrossing, and Amazon-Music.}
\zm{ML-100K\footnote{\url{https://grouplens.org/datasets/movielens/}} contains 100,000 user-item ratings and serves as a lightweight benchmark for evaluating model performance on smaller datasets.}
ML-1M includes over one million ratings and provides a more realistic testbed for assessing the scalability of RecRankerEval across larger interaction volumes. 
For BookCrossing, we follow the preprocessing 
\zym{setups}
in the RecRanker paper\zym{~\cite{luo2023recranker}} by extracting user ratings on a 1–10 scale and applying a 10-core filtering strategy~\cite{wang2019kgat}
\zym{to control data sparsity.}
\zym{\qq{Since} BookCrossing does not contain timestamp information, we follow RecRanker’s strategy of simulating timestamps by assigning them randomly, which enables temporal data partitioning during evaluation. }
\qq{Moreover, because} RecRanker does not release the processed dataset or 
\zym{preprocessing scripts,}
we re-implement the \zym{dataset} preprocessing pipeline to ensure consistency with the original settings. 
\zm{To further assess the generalisability of RecRankerEval across domains, we include the 
\zym{additional}
Amazon-Music dataset in our experiments.}
We 
\zym{consider}
each music title as an individual item and use the corresponding user ratings (on a 1–5 scale) as interactions.
\zym{We apply the same \qq{aforementioned} 10-core filtering strategy to ensure \qq{a} sufficient interaction density for training and evaluation.}
The 
statistics of \qq{the}
\zym{used} datasets are presented in Table~\ref{tab2}.\looseness -1

\vspace{-2mm}
\subsection{\zym{Baselines}}\label{s5.2}
\zym{We} report the performance of \qq{the} initial recommendation models as baselines for evaluating the effectiveness of RecRanker.
To reproduce the results \qq{of RecRanker's original paper}, we leverage MF~\cite{koren2009matrix} and LightGCN~\cite{he2020lightgcn}, \qq{both of} which \qq{were}  
originally used in RecRanker. 
\qq{We also} incorporate 
\zm{another}
state-of-the-art model, XSimGCL~\cite{yu2023xsimgcl}, to 
\qq{further} investigate the impact of \qq{the} initial recommendation models on \qq{the} recommendation performance. \qq{We describe below these models}: \looseness -1

\noindent $\bullet$  \textbf{MF~\cite{koren2009matrix}}: 
Matrix Factorisation (MF) is a 
\zm{well-established} collaborative filtering model that decomposes the user-item interaction matrix into low-dimensional matrices, capturing user preferences and the latent features of items;

\noindent $\bullet$  \textbf{LightGCN~\cite{he2020lightgcn}}:
\zym{LightGCN is a graph-based collaborative filtering model that simplifies prior GNN-based recommender architectures (e.g., NGCF~\cite{wang2019neural}) by removing feature transformations and non-linear activations;} 

\noindent $\bullet$  \textbf{XSimGCL~\cite{yu2023xsimgcl}}:
XSimGCL 
\zm{leverages}
noise-based embedding augmentation and contrastive learning to improve the consistency of node representations in user-item interaction graph, 
enabling a more effective capture of user interaction similarities. 
\pageenlarge{3}

\subsection{Evaluation Protocol and Implementation Details}\label{s5.3}
\pageenlarge{1}
We adopt the same evaluation metrics as used in the original RecRanker paper~\cite{luo2023recranker}, namely Hit Ratio (HR) and Normalised Discounted Cumulative Gain (NDCG), to measure top-$k$ recommendation performance.
Following common practice in recent LLM-based recommendation work~\cite{zhang2023recommendation,luo2023recranker}, we set $k$ to 3 and 5, due to the input token length limitation of the LLMs. 
\zym{As \qq{mentioned} in Section~\ref{s4.4}, we use Llama-2 (7B)\cite{touvron2023llama} to reproduce RecRanker's original results, and additionally include Llama-3.1-8B-Instruct\cite{dubey2024llama} and GPT-3.5 Turbo~\cite{ye2023comprehensive} for \qq{an} extended evaluation.}
\zym{To reduce training cost, we train all LLMs using a single NVIDIA RTX A6000 48GB GPU, in contrast to RecRanker’s original setup of 16 A800 GPUs. 
We \qq{also} apply LoRA~\cite{hu2021lora} for efficient fine-tuning. }
For \qq{a} fair comparison, we follow RecRanker's hyperparameter configurations
\zym{for both Llama2 and Llama3}: a learning rate of 2e-5, a context length of 2048, batch size of 2, and gradient accumulation steps of 64. 
We train the models using a cosine learning rate scheduler with 50 warm-up steps.
\zym{During the inference stage, we follow RecRanker's paper \qq{and} use the vLLM~\cite{kwon2023efficient} framework, setting}
the temperature to 0.1, top-$k$ sampling to 40, and nucleus sampling (top-$p$) to 0.1. 

\subsection{RQ1: Reproducing RecRanker}\label{s5.4}
\begin{table}
    \centering
    \renewcommand{\arraystretch}{0.85}
    \setlength{\tabcolsep}{4pt}
    \caption{\zym{\qq{Results} of \qq{the} RecRanker variants with MF and LightGCN on ML-100K and ML-1M. The superscript $^*$ indicates statistically significant improvements over the \qq{baseline} model using Holm-Bonferroni corrected paired t-test (p < 0.05).} }
    \resizebox{1\linewidth}{!}{
    \begin{tabular}{lcccccccc}
        \toprule
        \multirow{2}{*}{Method} & \multicolumn{4}{c}{ML-100K} & \multicolumn{4}{c}{ML-1M} \\
        \cmidrule(lr){2-5} \cmidrule(lr){6-9} 
        & H@3 $\uparrow$ & N@3 $\uparrow$ & H@5 $\uparrow$ & N@5 $\uparrow$ 
        & H@3 $\uparrow$ & N@3 $\uparrow$ & H@5 $\uparrow$ & N@5 $\uparrow$ 
         \\
        \midrule
        \zym{MF (Base)}                     & {0.0421}$^*$  & {0.0309}$^*$ & {0.0690}$^*$ & {0.0419}$^*$ & {0.0250}$^*$ & {0.0185}$^*$ & {0.0403}$^*$  & {0.0248}$^*$  \\
        $RecRanker_{pointwise}^{MF}$  & \underline{0.0807}$^*$  & \underline{0.0729}$^*$ & \underline{0.0966}$^*$ & \underline{0.0795}$^*$ & \textbf{0.0540} & \textbf{0.0511} & \textbf{0.0598} & \textbf{0.0534}  \\
        $RecRanker_{pairwise}^{MF}$   & {0.0499}$^*$  & {0.0346}$^*$ & {0.0701}$^*$ & {0.0430}$^*$ & {0.0237}$^*$ & {0.0168}$^*$ & {0.0373}$^*$ & {0.0223}$^*$ \\
        $RecRanker_{listwise}^{MF}$   & {0.0475}$^*$  & {0.0360}$^*$ & {0.0729}$^*$ & {0.0463}$^*$ & \underline{0.0257}$^*$ & {0.0188}$^*$ & \underline{0.0418}$^*$ & {0.0254}$^*$  \\
        $RecRanker_{hybrid}^{MF}$     & \textbf{0.0851} & \textbf{0.0765} & \textbf{0.1017} & \textbf{0.0833} & \textbf{0.0540} & \underline{0.0510} & \textbf{0.0598} & \underline{0.0533}  \\

        \midrule
        \midrule
        \zym{LightGCN (Base)}                     & {0.0475}$^*$ & {0.0335}$^*$ & {0.0712}$^*$ & {0.0433}$^*$ & {0.0270}$^*$ & {0.0194}$^*$ & {0.0428}$^*$ & {0.0258}$^*$  \\
        $RecRanker_{pointwise}^{LightGCN}$  & \textbf{0.0839} & \underline{0.0746} & \textbf{0.0966} & 0.0800 & \underline{0.0593} & \underline{0.0557} & \textbf{0.0659} & \textbf{0.0583}  \\
        $RecRanker_{pairwise}^{LightGCN}$   & {0.0499}$^*$ & {0.0351}$^*$ & \underline{0.0817}$^*$ & \underline{0.0482}$^*$ & {0.0270}$^*$ & {0.0195}$^*$ & {0.0436}$^*$ & {0.0263}$^*$ \\
        $RecRanker_{listwise}^{LightGCN}$   & \underline{0.0510}$^*$ & {0.0350}$^*$ & {0.0764}$^*$ & {0.0454}$^*$ & {0.0295}$^*$ & {0.0215}$^*$ & {0.0456}$^*$ & \underline{0.0281}$^*$  \\
        $RecRanker_{hybrid}^{LightGCN}$      & \textbf{0.0839} & \textbf{0.0748} & \textbf{0.0966} & \textbf{0.0802} & \textbf{0.0598} & \textbf{0.0559} & \underline{0.0658} & \textbf{0.0583}  \\
        \bottomrule
    \end{tabular}
    }
\label{tab-3}
\end{table} 
\begin{sloppypar}
Table~\ref{tab-3} reports the results of reproducing RecRanker on the two original datasets (ML-100K and ML-1M) using two initial recommendation models (MF and LightGCN). 
We evaluate four instruction tuning variants of RecRanker
--$RecRanker_{pointwise}$, $RecRanker_{pairwise}$, $RecRanker_{listwise}$, and $RecRanker_{hybrid}$--each corresponding to a specific prompt design and tuning strategy described in Section~\ref{s3.4}.
\zym{For clarity, we denote each variant with a superscript indicating the initial model used, such as $RecRanker_{pointwise}^{MF}$ and $RecRanker_{listwise}^{LightGCN}$.}
\zym{Our reproduced results show clear performance gains over the base models across all configurations, confirming the \qq{effectiveness} of RecRanker’s design.}
Specifically, $RecRanker_{hybrid}$ achieves the best overall performance in most settings, while $RecRanker_{pointwise}$ consistently outperforms both $RecRanker_{pairwise}$ and $RecRanker_{listwise}$. 
\qq{Moreover}, $RecRanker_{listwise}$ performs better than $RecRanker_{pairwise}$ across all configurations. 
These findings show that our RecRankerEval framework effectively reproduces RecRanker’s original results and provides a stable foundation for further analysis.
\end{sloppypar}
\begin{table*}
    \centering
    \renewcommand{\arraystretch}{0.85}
    \setlength{\tabcolsep}{4pt}
    \caption{\zym{Performance of \qq{the} RecRanker variants after removing prompt leakage, across multiple datasets and initial recommendation models. The superscript $^*$ indicates statistically significant differences compared to the respective base model (Holm-Bonferroni corrected paired t-test, p < 0.05).}}
    \vspace{-3mm}
    \resizebox{0.8\linewidth}{!}{
    \begin{tabular}{lcccccccccccccccc}
        \toprule
        \multirow{2}{*}{Method} & \multicolumn{4}{c}{ML-100K} & \multicolumn{4}{c}{ML-1M} &\multicolumn{4}{c}{BookCrossing} & \multicolumn{4}{c}{Amazon-Music}\\
        \cmidrule(lr){2-5} \cmidrule(lr){6-9} \cmidrule(lr){10-13} \cmidrule(lr){14-17}
        & H@3 $\uparrow$ & N@3 $\uparrow$ & H@5 $\uparrow$ & N@5 $\uparrow$ 
        & H@3 $\uparrow$ & N@3 $\uparrow$ & H@5 $\uparrow$ & N@5 $\uparrow$ 
        & H@3 $\uparrow$ & N@3 $\uparrow$ & H@5 $\uparrow$ & N@5 $\uparrow$
        & H@3 $\uparrow$ & N@3 $\uparrow$ & H@5 $\uparrow$ & N@5 $\uparrow$
         \\
        \midrule

        \zym{MF (Base)}         & {0.0421}$^*$  & {0.0309}$^*$ & {0.0690}$^*$ & {0.0419}$^*$ & {0.0250} & {0.0185} & \underline{0.0403}  & {0.0248} &\textbf{0.0269} &\textbf{0.0194} &\textbf{0.0358} &\textbf{0.0230}  &\underline{0.0266}  &\underline{0.0190}  &\textbf{0.0430}  & \underline{0.0256}             \\
        $RecRanker_{pointwiseFix}^{MF}$   &\underline{0.0488}  &\underline{0.0377}  &\underline{0.0701}  &\textbf{0.0466} & \underline{0.0257}&\textbf{0.0195} &{0.0391} &{0.0250} &{0.0180}$^*$ &{0.0128}$^*$ &{0.0242}$^*$ &{0.0154}$^*$  &{0.0198}$^*$ &{0.0140}$^*$  &{0.0379}$^*$  &{0.0214}$^*$\\
        $RecRanker_{pairwise}^{MF}$ & \textbf{0.0499}  & {0.0346}$^*$ & \underline{0.0701} & {0.0430}$^*$ & {0.0237}$^*$ & {0.0168}$^*$ & {0.0373}$^*$ & {0.0223}$^*$&{0.0141}$^*$ &{0.0110}$^*$ &{0.0219}$^*$ &{0.0142}$^*$ &{0.0207}$^*$  &{0.0139}$^*$  &{0.0353}$^*$  &{0.0198}$^*$ \\
        $RecRanker_{listwise}^{MF}$    & {0.0475}  & {0.0360} & \textbf{0.0729} & {0.0463} & \underline{0.0257} & \underline{0.0188} & \textbf{0.0418} & \textbf{0.0254} &\underline{0.0214}$^*$ &\underline{0.0156}$^*$ &\underline{0.0264}$^*$ &\underline{0.0176}$^*$ &\textbf{0.0275}  &\textbf{0.0202}  &\underline{0.0422}  &\textbf{0.0261} \\
        $RecRanker_{hybridFix}^{MF}$  &\underline{0.0488}  &\textbf{0.0376}  &{0.0700}  &\underline{0.0465}  & \textbf{0.0258}&\textbf{0.0195} &{0.0389} &\underline{0.0249}&{0.0174}$^*$ &{0.0126}$^*$ &{0.0242}$^*$ &{0.0154}$^*$ &{0.0198}$^*$  &{0.0141}$^*$  &{0.0370}$^*$  &{0.0213}$^*$    \\

        \midrule
        \midrule
        \multirow{2}{*}{Method} & \multicolumn{4}{c}{ML-100K} & \multicolumn{4}{c}{ML-1M} &\multicolumn{4}{c}{BookCrossing} & \multicolumn{4}{c}{Amazon-Music} \\
        \cmidrule(lr){2-5} \cmidrule(lr){6-9} \cmidrule(lr){10-13} \cmidrule(lr){14-17}
        & H@3 $\uparrow$ & N@3 $\uparrow$ & H@5 $\uparrow$ & N@5 $\uparrow$ 
        & H@3 $\uparrow$ & N@3 $\uparrow$ & H@5 $\uparrow$ & N@5 $\uparrow$ 
        & H@3 $\uparrow$ & N@3 $\uparrow$ & H@5 $\uparrow$ & N@5 $\uparrow$ 
        & H@3 $\uparrow$ & N@3 $\uparrow$ & H@5 $\uparrow$ & N@5 $\uparrow$
         \\
        \midrule
        \zym{LightGCN (Base)}  & {0.0475}$^*$ & {0.0335}$^*$ & {0.0712}$^*$ & {0.0433}$^*$ & \underline{0.0270}$^*$ & {0.0194}$^*$ & {0.0428}$^*$ & {0.0258}$^*$   & \textbf{0.0317} & \textbf{0.0236} & \textbf{0.0414} & \textbf{0.0276} & \underline{0.0380} & \underline{0.0277} & \textbf{0.0568} & \underline{0.0346}                   \\
        $RecRanker_{pointwiseFix}^{LightGCN}$    & \textbf{0.0510} & \underline{0.0365} & {0.0722}$^*$ & 0.0453 & {0.0240}$^*$ & {0.0172}$^*$ & {0.0423}$^*$ & {0.0247}$^*$  & {0.0236}$^*$ & {0.0169}$^*$ & {0.0326}$^*$ & {0.0207}$^*$ & {0.0327}$^*$ & {0.0248}$^*$ & {0.0482}$^*$ & {0.0312}$^*$ \\
        $RecRanker_{pairwise}^{LightGCN}$    & \underline{0.0499}$^*$ & {0.0351} & \textbf{0.0817} & \textbf{0.0482} & \underline{0.0270}$^*$ & \underline{0.0195}$^*$ & \underline{0.0436}$^*$ & \underline{0.0263}$^*$  & {0.0186}$^*$ & {0.0140}$^*$ & {0.0293}$^*$ & {0.0183}$^*$ & {0.0301}$^*$ & {0.0209}$^*$ & {0.0473}$^*$ & {0.0282}$^*$\\
        $RecRanker_{listwise}^{LightGCN}$    & \textbf{0.0510} & {0.0350} & \underline{0.0764} & 0.0454 & \textbf{0.0295} & \textbf{0.0215} & \textbf{0.0456} & \textbf{0.0281}   & \underline{0.0270}$^*$ & \underline{0.0204}$^*$ & \underline{0.0338}$^*$ & \underline{0.0233}$^*$ & \textbf{0.0391} & \textbf{0.0294} & \underline{0.0547} & \textbf{0.0358}\\
        $RecRanker_{hybridFix}^{LightGCN}$    & \textbf{0.0510} & \textbf{0.0370} & {0.0722}$^*$ & \textbf{0.0458} & {0.0244}$^*$ & {0.0175}$^*$ & {0.0428}$^*$ & {0.0251}$^*$   & {0.0242}$^*$ & {0.0175}$^*$ & {0.0326}$^*$ & {0.0210}$^*$ & {0.0336}$^*$ & {0.0257}$^*$ & {0.0465}$^*$ & {0.0309}$^*$  \\
        
        \midrule
        \midrule
        \multirow{2}{*}{Method} & \multicolumn{4}{c}{ML-100K} & \multicolumn{4}{c}{ML-1M} &\multicolumn{4}{c}{BookCrossing} & \multicolumn{4}{c}{Amazon-Music} \\
        \cmidrule(lr){2-5} \cmidrule(lr){6-9} \cmidrule(lr){10-13} \cmidrule(lr){14-17}
        & H@3 $\uparrow$ & N@3 $\uparrow$ & H@5 $\uparrow$ & N@5 $\uparrow$ 
        & H@3 $\uparrow$ & N@3 $\uparrow$ & H@5 $\uparrow$ & N@5 $\uparrow$ 
        & H@3 $\uparrow$ & N@3 $\uparrow$ & H@5 $\uparrow$ & N@5 $\uparrow$ 
        & H@3 $\uparrow$ & N@3 $\uparrow$ & H@5 $\uparrow$ & N@5 $\uparrow$
         \\
        \midrule
        \zym{XSimGCL (Base)}         & {0.0496}$^*$ & {0.0369}$^*$ & {0.0788}$^*$ & {0.0489}$^*$ & \underline{0.0292}$^*$ & \underline{0.0214}$^*$ & \underline{0.0443}$^*$ & \underline{0.0275}$^*$  &\textbf{0.0393} &\textbf{0.0308} &\textbf{0.0503} &\textbf{0.0353}  &\underline{0.0440}  &\underline{0.0325}  &\underline{0.0568}  &  \underline{0.0378}             \\
        $RecRanker_{pointwiseFix}^{XSimGCL}$ &\underline{0.0520}$^*$  &{0.0400}$^*$  &\underline{0.0817}$^*$  &{0.0522}$^*$ &{0.0270}$^*$  &{0.0200}$^*$  &{0.0438}$^*$  &{0.0270}$^*$ &{0.0253}$^*$ &{0.0190}$^*$ &\underline{0.0366}$^*$ &{0.0236}$^*$&{0.0353}$^*$  &{0.0249}$^*$  &{0.0516}$^*$  &{0.0318}$^*$ \\
        $RecRanker_{pairwise}^{XSimGCL}$  & {0.0361}$^*$ & {0.0248}$^*$ & {0.0584}$^*$ & {0.0340}$^*$ & {0.0255}$^*$ & {0.0182}$^*$ & {0.0418}$^*$ & {0.0249}$^*$ &{0.0152}$^*$ &{0.0104}$^*$ &{0.0281}$^*$ &{0.0156}$^*$ &{0.0240}$^*$  &{0.0187}$^*$  &{0.0499}$^*$  &{0.0294}$^*$ \\
        $RecRanker_{listwise}^{XSimGCL}$  & \textbf{0.0585} & \textbf{0.0424} & \textbf{0.0943} & \textbf{0.0571} & \textbf{0.0303} & \textbf{0.0222} & \textbf{0.0461} & \textbf{0.0286} &\underline{0.0284}$^*$ &\underline{0.0218}$^*$ &{0.0358}$^*$ &\underline{0.0248}$^*$ &\textbf{0.0443}  &\textbf{0.0328}  &\textbf{0.0584}  &\textbf{0.0386}  \\
        $RecRanker_{hybridFix}^{XSimGCL}$   &\underline{0.0520}$^*$  &\underline{0.0404}$^*$  &\underline{0.0817}$^*$  &\underline{0.0526}$^*$ &{0.0265}$^*$  &{0.0196}$^*$  &{0.0428}$^*$  &{0.0262}$^*$ &{0.0253} &{0.0189} &{0.0360} &{0.0233} &{0.0318}$^*$  &{0.0226}$^*$  &{0.0451}$^*$  &{0.0281}$^*$  \\
        \bottomrule
    \end{tabular}
    }
\label{tab-4}
\vspace{-3mm}
\end{table*}
However, we observe that $RecRanker_{pointwise}$ achieves \qq{an} abnormally high 
performance across all datasets, significantly outperforming the $RecRanker_{pairwise}$ and $RecRanker_{listwise}$.
To investigate this result, we re-examine the prompt construction process described in \zym{Section~\ref{s3.3}}, particularly focusing on the use of hint scores in pointwise prompts. 
We find that in the original design, \qq{the} pointwise prompts include explicit scores derived from user ratings: interacted (i.e., ground-truth) items are assigned high scores, while non-interacted items receive low or zero scores. 
This introduces a risk of data leakage, as the model effectively receives direct 
\zym{signals}
about the correct answers within the input prompt. 
\zym{Such leakage can artificially boost the model’s performance by reducing the learning burden and enabling shortcut-based predictions.}
To verify this hypothesis, we construct a corrected variant, $RecRanker_{pointwiseFix}$, by removing the hint scores from \qq{the} pointwise prompts.
\zym{Table~\ref{tab-4} presents the updated results across four datasets \qq{and} three initial recommendation models (MF, LightGCN, XSimGCL).}
\zym{Compared to $RecRanker_{pointwise}$, the performance of $RecRanker_{pointwiseFix}$ drops significantly, aligning more closely with $RecRanker_{pairwise}$ and $RecRanker_{listwise}$.}
\zym{In this corrected setting, $RecRanker_{listwise}$ emerges as the most effective variant across most datasets and initial recommendation models.}
\zym{These findings show that the original pointwise variant’s strong performance was likely the result of prompt leakage, and suggest the need for careful prompt design when evaluating instruction tuning strategies.}\looseness-1

\pageenlarge{4}
To further investigate the generalisability of RecRankerEval, Table~\ref{tab-4} also includes results on two additional datasets \zym{in relation to the original RecRanker paper}: BookCrossing and Amazon-Music. 
\zym{On BookCrossing, all RecRanker variants underperform relative to their respective initial recommendation model baselines.}
Although we follow the original paper’s preprocessing procedure, the lack of 
\zym{actual} timestamps requires \qq{the simulation of} interaction times for sequential partitioning, which may introduce noise and hinder training. 
\zym{In contrast, Amazon-Music provides real interaction timestamps, allowing for a more realistic sequential split.}
\zym{On this dataset, \qq{the} RecRanker variants deliver more competitive results, with $RecRanker_{listwise}$ achieving the best overall performance and significantly outperforming both the baseline and other variants.}
These results suggest that 
\zym{RecRanker}
can generalise to new domains, but the quality of temporal and interaction data plays a critical role in enabling effective LLM-based training and evaluation.\looseness -1

\subsection{RQ2: Impact of Different Initial Recommendation Models}\label{s5.5}
To investigate the impact of different initial recommendation models on the performance of 
\zym{RecRanker,}
we conduct experiments using three representative models: MF and LightGCN, both originally used in RecRanker, and XSimGCL, a state-of-the-art contrastive learning-based model. 
\zym{Table~\ref{tab-4} reports the results across four datasets, covering all instruction tuning variants and base models.}
\zym{Across all three initial recommendation models, }
we observe consistent trends across \qq{the} datasets.
$RecRanker_{listwise}$ and $RecRanker_{hybridFix}$ generally outperform $RecRanker_{pairwise}$ and $RecRanker_{pointwiseFix}$, with $RecRanker_{listwise}$ achieving the best performance in most configurations. 
This effect is particularly pronounced under XSimGCL, where $RecRanker_{listwise}$ achieves the highest scores across all four datasets, such as an H@3 of 0.0585 on ML-100K and 0.0443 on Amazon-Music. 
This suggests that when the initial recommendation model provides \qq{a} higher quality initial ranking list (as XSimGCL does), the LLM benefits more from \qq{the} instruction tuning strategies that leverage full ranking contexts.
Comparing \qq{results} across \qq{the} initial recommendation models, we find that stronger initial recommendation models, such as XSimGCL, lead to \qq{a}
\zym{better overall performance for all tuning variants.}
For example, on Amazon-Music, $RecRanker_{listwise}$ built on XSimGCL achieves \qq{an} H@3 of 0.0443, significantly higher than its MF (0.0275) and LightGCN (0.0391) variants.
However, we observe that \qq{the} RecRanker variants based on classical initial recommendation models (e.g., MF) benefit more significantly from instruction tuning, suggesting that LLM-based approaches can effectively enhance recommendation performance when the base model provides \qq{a} limited ranking quality.
In summary, our findings demonstrate that both the choice of initial recommendation model and the design of the instruction tuning strategy jointly affect the final recommendation quality.
Stronger initial recommendation models benefit more from expressive prompt structures such as $listwise$ and $hybrid$, while weaker \zym{base} models gain more from the LLM’s ability to refine coarse initial rankings through fine-tuning.
\pageenlarge{3}


\begin{table*}
    \centering
    \renewcommand{\arraystretch}{0.85}
    \setlength{\tabcolsep}{4pt}
    \caption{\zym{Impact of \qq{the} user sampling strategies on \qq{performance} across \qq{various tuning variants and datasets}. The superscript $^*$ indicates statistically significant differences based on the Holm-Bonferroni corrected paired t-test with $p$< 0.05.}\looseness -1}
    \resizebox{0.7\linewidth}{!}{
    \begin{tabular}{lcccccccccccc}
        \toprule
        \multirow{2}{*}{Method} & \multicolumn{4}{c}{ML-100K} & \multicolumn{4}{c}{ML-1M} & \multicolumn{4}{c}{Amazon-Music}\\
        \cmidrule(lr){2-5} \cmidrule(lr){6-9}  \cmidrule(lr){10-13}
        & H@3 $\uparrow$ & N@3 $\uparrow$ & H@5 $\uparrow$ & N@5 $\uparrow$ 
        & H@3 $\uparrow$ & N@3 $\uparrow$ & H@5 $\uparrow$ & N@5 $\uparrow$ 
        & H@3 $\uparrow$ & N@3 $\uparrow$ & H@5 $\uparrow$ & N@5 $\uparrow$\\
        \midrule
        \zym{LightGCN (Base)}                     & {0.0475}$^*$ & {0.0335}$^*$ & {0.0712}$^*$ & {0.0433}$^*$ & \underline{0.0270}$^*$ & \underline{0.0194}$^*$ & \underline{0.0428}$^*$ & \underline{0.0258}$^*$ & \underline{0.0380} & \underline{0.0277} & \textbf{0.0568} & \underline{0.0346} \\
        Random$_{pointwiseFix}$  & \underline{0.0510} & \underline{0.0358} & {0.0732}$^*$ & 0.0450 & {0.0234}$^*$ & {0.0169}$^*$ & {0.0416}$^*$ & {0.0244}$^*$  & {0.0310}$^*$ & {0.0240}$^*$ & {0.0456}$^*$ & {0.0300}$^*$\\
        Random$_{pairwise}$   & {0.0499}$^*$ & 0.0342 & \textbf{0.0817} & \textbf{0.0474} & {0.0220}$^*$ & {0.0157}$^*$ & {0.0413}$^*$ & {0.0235}$^*$  & {0.0250}$^*$ & {0.0185}$^*$ & {0.0439}$^*$ & {0.0264}$^*$\\
        Random$_{listwise}$   & \textbf{0.0520} & \textbf{0.0362} & \underline{0.0775} & \underline{0.0466} & \textbf{0.0287} & \textbf{0.0209} & \textbf{0.0442} & \textbf{0.0272} & \textbf{0.0382} & \textbf{0.0294} & \underline{0.0547} & \textbf{0.0360}\\
        Random$_{hybridFix}$     & {0.0499}$^*$ & 0.0353 & {0.0732}$^*$ & 0.0450 & {0.0237}$^*$ & {0.0171}$^*$ & {0.0416}$^*$ & {0.0244}$^*$ & {0.0336}$^*$ & {0.0257}$^*$ & {0.0465}$^*$ & {0.0309}$^*$\\
        \midrule
        \zym{LightGCN (Base)}                     & {0.0475}$^*$ & {0.0335}$^*$ & {0.0712}$^*$ & {0.0433}$^*$ & \underline{0.0270}$^*$ & \underline{0.0194}$^*$ & {0.0428}$^*$ & \underline{0.0258}$^*$ &\underline{0.0380} & \underline{0.0277} & \underline{0.0568} & \underline{0.0346} \\
        DBSCAN$_{pointwiseFix}$  & \underline{0.0510} & \underline{0.0363} & {0.0722}$^*$ & {0.0452}$^*$ & {0.0242}$^*$ & {0.0174}$^*$ & {0.0424}$^*$ & {0.0249}$^*$ &{0.0310}$^*$ &{0.0240}$^*$ &{0.0473}$^*$ &{0.0308}$^*$\\
        DBSCAN$_{pairwise}$   & {0.0499}$^*$ & {0.0347}$^*$ & \textbf{0.0807} & \textbf{0.0474} & {0.0260}$^*$ & {0.0182}$^*$ & \underline{0.0432}$^*$ & {0.0253}$^*$ &{0.0275}$^*$ &{0.0211}$^*$ &{0.0456}$^*$ &{0.0286}$^*$ \\
        DBSCAN$_{listwise}$   & {0.0499}$^*$ & 0.0355 & \underline{0.0743}$^*$ & 0.0456 & \textbf{0.0303} & \textbf{0.0222} & \textbf{0.0462} & \textbf{0.0287} &\textbf{0.0391} &\textbf{0.0292} &\textbf{0.0591} &\textbf{0.0373} \\
        DBSCAN$_{hybridFix}$  & \textbf{0.0520} & \textbf{0.0375} & {0.0722}$^*$ & \underline{0.0459} & {0.0245}$^*$ & {0.0177}$^*$ & {0.0423}$^*$ & {0.0250}$^*$  &{0.0318}$^*$ &{0.0248}$^*$ &{0.0491}$^*$ &{0.0318}$^*$\\
        \midrule
        \zym{LightGCN (Base)}                     & {0.0475}$^*$ & {0.0335}$^*$ & {0.0712}$^*$ & {0.0433}$^*$ & \underline{0.0270}$^*$ & {0.0194}$^*$ & {0.0428}$^*$ & {0.0258}$^*$  & \underline{0.0380} & \underline{0.0277} & \textbf{0.0568} & \underline{0.0346} \\
        KMeans$_{pointwiseFix}$  & \textbf{0.0510} & \underline{0.0365} & {0.0722}$^*$ & 0.0453 & {0.0240}$^*$ & {0.0172}$^*$ & {0.0423}$^*$ & {0.0247}$^*$ & {0.0327}$^*$ & {0.0248}$^*$ & {0.0482}$^*$ & {0.0312}$^*$\\
        KMeans$_{pairwise}$   & \underline{0.0499}$^*$ & {0.0351} & \textbf{0.0817} & \textbf{0.0482} & \underline{0.0270}$^*$ & \underline{0.0195}$^*$ & \underline{0.0436}$^*$ & \underline{0.0263}$^*$  & {0.0301}$^*$ & {0.0209}$^*$ & {0.0473}$^*$ & {0.0282}$^*$\\
        KMeans$_{listwise}$   & \textbf{0.0510} & {0.0350} & \underline{0.0764} & 0.0454 & \textbf{0.0295} & \textbf{0.0215} & \textbf{0.0456} & \textbf{0.0281}  & \textbf{0.0391} & \textbf{0.0294} & \underline{0.0547} & \textbf{0.0358}\\
        KMeans$_{hybridFix}$      & \textbf{0.0510} & \textbf{0.0370} & {0.0722}$^*$ & \textbf{0.0458} & {0.0244}$^*$ & {0.0175}$^*$ & {0.0428}$^*$ & {0.0251}$^*$ 
  & {0.0336}$^*$ & {0.0257}$^*$ & {0.0465}$^*$ & {0.0309}$^*$ \\
        \bottomrule
    \end{tabular}
    }
\label{tab-6}
\end{table*}
\subsection{RQ3: Impact of Different User Sampling Strategies}\label{s5.6}
\begin{sloppypar}
To evaluate the effect of different user sampling strategies on \qq{RecRanker’s} performance, we compare three approaches introduced in Section~\ref{s4.2}: KMeans (used in the original RecRanker \qq{paper}), DBSCAN (a density-based clustering method), and Random sampling.
Each sampling method is evaluated under four instruction tuning methods--$pointwiseFix$, $pairwise$, $listwise$, and $hybridFix$ \qq{(where we use the corrected pointwise prompts)}--achieving \qq{the} corresponding variants such as $Random_{listwise}$ and $DBSCAN_{pairwise}$.
Table~\ref{tab-6} reports the results across three datasets: ML-100K, ML-1M, and Amazon-Music, using LightGCN as the initial recommendation model. 
\zym{We adopt LightGCN to ensure consistent comparison with prior work and because of its stable performance across different datasets.}
\zym{We exclude \qq{the} BookCrossing \qq{dataset} due to its limited timestamp information, which may interfere with \qq{a} consistent sampling-based evaluation.}
\zym{We begin by comparing results across different sampling methods under the same tuning strategy.}
\zym{In the $listwise$ setting, $DBSCAN_{listwise}$ achieves the best results on all four metrics on \qq{the} ML-1M and Amazon-Music datasets, and performs competitively on ML-100K.} 
Similarly, in the $hybridFix$ setting, $DBSCAN_{hybridFix}$ outperforms both $KMeans_{hybridFix}$ and $Random_{hybridFix}$ on ML-100K and Amazon-Music, and 
\zym{\qq{achieves}}
the best perfrmance on ML-1M for N@5.
\zym{In the $pairwise$ setting, DBSCAN and KMeans achieve \qq{a} comparable performance overall: $KMeans_{pairwise}$ slightly outperforms on \qq{the} ML-100K and Amazon-Music datasets, whereas $DBSCAN_{pairwise}$ achieves the best results on ML-1M.}
\zym{Random sampling achieves \qq{a} strong performance under the $listwise$ variant—e.g., $Random_{listwise}$ achieves the best H@3 and N@3 on ML-100K and \qq{the} second best results on the other two datasets.}
\zym{However, in \qq{the} $pairwise$ and $hybridFix$ settings, \qq{the} Random variants consistently underperform, indicating that unstructured user sampling may be less effective when instruction tuning relies on finer-grained ranking signals.}
\zym{These results suggest that \qq{the} DBSCAN-based sampling generally provides more consistent improvements across different tuning methods.}
\zym{DBSCAN’s ability to group users based on local density and exclude outliers (labelled as noise) \qq{seems to} help produce more representative prompts.}
Although $Random_{listwise}$ performs competitively in some settings, 
\zym{Random sampling is generally less effective than DBSCAN and KMeans across most instruction tuning variants.}
\zym{In summary, \zym{we can conclude that }DBSCAN is the most effective sampling strategy across \qq{the used} datasets and tuning methods, while KMeans and Random show competitive but less consistent \qq{performances}.}
\end{sloppypar}

\begin{figure}
\centering
\includegraphics[width=1\linewidth]{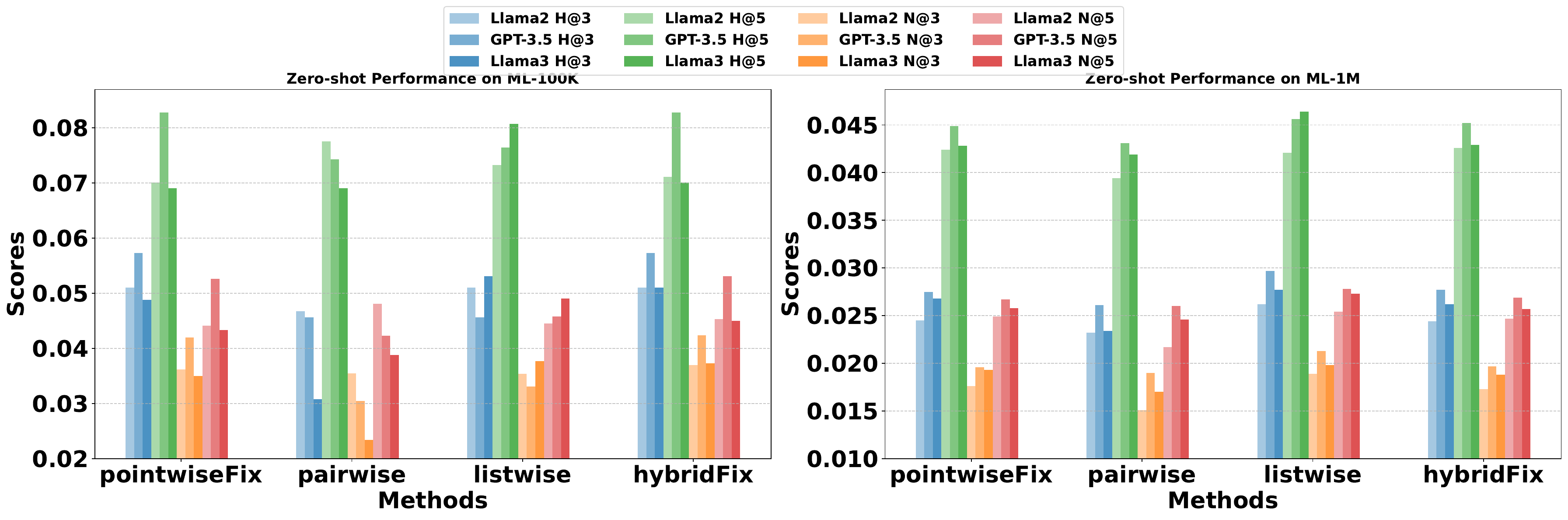}
\caption{Zero-shot performance comparison of Llama2, GPT-3.5, and Llama3 on top-$k$ recommendation} 
 
\label{fig2}
\end{figure} 
\begin{figure}
\centering
\includegraphics[width=1\linewidth]{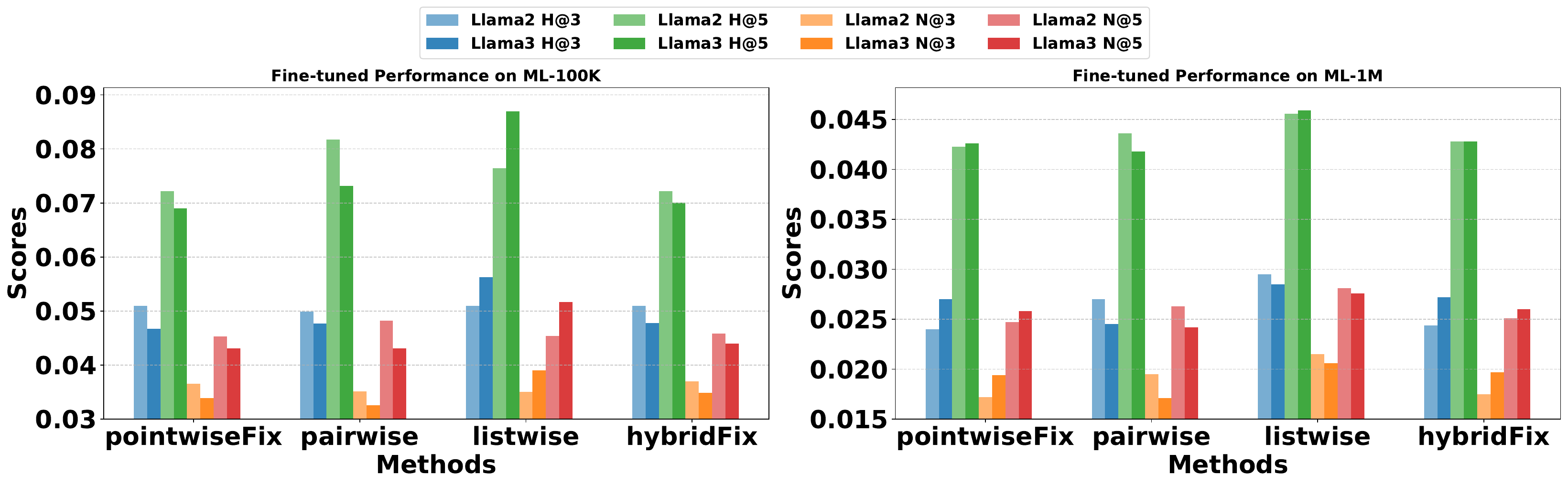}
\caption{Instruction-tuned performance comparison of Llama2 and Llama3 on top-$k$ recommendation}
 
\label{fig3}
\end{figure}
\subsection{RQ4: Impact of the Different LLMs}\label{s5.7}
\pageenlarge{3}
To investigate how the choice of large language models (LLMs) affects \qq{the} recommendation performance,
\zym{we compare \qq{RecRanker} using three LLMs: Llama2, Llama3, and GPT-3.5.}
\zym{Llama2 and Llama3 are open models that support instruction tuning and are evaluated under both \qq{the} zero-shot and fine-tuned settings.}
\zym{GPT-3.5, a closed model accessed via API, is evaluated only in the zero-shot setting due to the lack of fine-tuning support.}
\zym{Figures~\ref{fig2} and~\ref{fig3} present the performance results on ML-100K and ML-1M 
under the zero-shot and instruction-tuned settings, respectively.}
\zym{Given the space constraints, we focus on these two datasets to ensure an alignment with the original RecRanker paper.}
In both figures, each group of bars represents one instruction tuning variant\zym{--}$pointwiseFix$, $pairwise$, $listwise$, and $hybridFix$\zym{--}while 
\zym{each colour represents a particular evaluation metric: H@3 (blue), H@5 (green), N@3 (orange), and N@5 (red).}
\pageenlarge{3}

\zym{Figure~\ref{fig2} shows the performance of all three LLMs under the zero-shot setting, evaluated across the four instruction tuning variants. 
Overall, no single model consistently dominates across all variants and metrics.}
\zym{GPT-3.5 achieves the best overall performance under \qq{the} $pointwiseFix$ and $hybridFix$ variants, particularly on ML-1M, where it obtains the highest N@3 and N@5 \qq{scores}.}
In contrast, Llama3 performs better under \qq{the} $listwise$ variant, especially on ML-100K, 
\zym{where it achieves the best results in all 4 metric combinations.}
\zym{Llama2 achieves \qq{a} competitive performance in certain cases—-such as achieving comparable H@3 \qq{scores} under \qq{the} $pointwiseFix$ variant on ML-100K—-but generally underperforms relative to GPT-3.5 and Llama3 in 23 out of 32 \qq{cases}.}
These results indicate that GPT-3.5 and Llama3 generally deliver \qq{a} stronger zero-shot performance compared to Llama2, 
\zym{especially when the instruction tuning variant and dataset characteristics align well with each model’s capabilities.}

\zym{\qq{Furthermore, as} shown in Figure~\ref{fig3}, both Llama2 and Llama3 exhibit notable performance improvements after instruction tuning, compared to their zero-shot variants in Figure~\ref{fig2}.}
This 
\zym{improvement}
is consistent across all instruction tuning variants ($pointwiseFix$, $pairwise$, $listwise$, $hybridFix$) and evaluation metrics on both datasets.
\zym{Moreover, Llama3 consistently achieves the best performance, particularly under the $listwise$ variant, where it obtains the best H@3and N@3 \qq{performances} on ML-100K.}
Llama3 also maintains strong results under $hybridFix$, suggesting that it is 
\zym{better adapted to \qq{the} top-$k$ ranking task} after tuning. 
\zym{\qq{In contrast,} Llama2, while showing clear gains from fine-tuning, generally remains less effective than Llama3 in most configurations.}
\zym{These results confirm the effectiveness of instruction tuning in enhancing \qq{the} top-$k$ recommendation performance. 
Furthermore, 
they 
\zym{the results}
show the importance of selecting capable backbone models: models like Llama3 are more effective in leveraging the tuning data and instruction variants, making them more suitable for future LLM-based top-$k$ recommendation tasks.}

\pageenlarge{3}
\section{Conclusions}\label{s6}
In this paper, we \qq{proposed} RecRankerEval, a reproducible and extensible framework
\zym{built upon RecRanker,}
designed to support the systematic evaluation of \qq{RecRanker's components as well as} LLM-based top-$k$ recommender systems \qq{that follow a similar paradigm}. 
Our experimental results show that \qq{we could successfully reproduce} the main results reported in the original RecRanker, \qq{while also evaluating the impact of different fine-tuning, user sampling strategies, initial recommendation models and backbone LLMs on the overall performance of the model} .
\zym{\qq{We also identified and corrected}}
a data leakage issue in the pointwise method caused by prompts containing ground-truth information.
\zym{Beyond reproduction, RecRankerEval \qq{indeed} enables comparative analysis across model design choices. 
We \qq{observed} that stronger initial recommendation models and more capable LLMs (e.g., Llama3) achieve consistent performance improvements. 
Among \qq{the} instruction tuning methods, the listwise variant achieves the most effective results across datasets.}
\qq{Furthermore}, we \qq{found} that user sampling strategies influence \qq{the} recommendation performance, with methods like DBSCAN performing better than random and 
\zym{KMeans-based approaches.}
\zm{Overall, RecRankerEval offers a practical framework for reproducible experimentation and systematic analysis of LLM-based top-$k$ recommendation models across key design choices.}

\clearpage
\balance

\bibliographystyle{ACM-Reference-Format}
\bibliography{reference}
\end{document}